\documentstyle[NATO,epsf]{crckapb} 

\def\pbarp{$\overline{p}p $}            
\def\ppbar{$p\overline{p} $}            
\def\ttbar{$t\overline{t}$}             
\def\bbbar{$b\overline{b}$}             
\def\D0{D\O}                            
\def\bbar{$\overline{b}$}               
\def\lumunits{cm$^{-2}$s$^{-1}$}        
\def\met{\mbox{${\hbox{$E$\kern-0.6em\lower-.1ex\hbox{/}}}_T$ }} 
\def\d0draft{}

\def\err#1#2#3 {{\it Erratum} {\bf#1},{\ #2} (19#3)}
\def\ib#1#2#3 {{\it ibid.} {\bf#1},{\ #2} (19#3)}
\def\nc#1#2#3 {Nuovo Cim. {\bf#1} ,#2(19#3)}
\def\nim#1#2#3 {Nucl. Instr. Meth. {\bf#1},{\ #2} (19#3)}
\def\np#1#2#3 {Nucl. Phys. {\bf#1},{\ #2} (19#3)}
\def\pl#1#2#3 {Phys. Lett. {\bf#1},{\ #2} (19#3)}
\def\prev#1#2#3 {Phys. Rev. {\bf#1},{\ #2} (19#3)}
\def\prl#1#2#3 {Phys. Rev. Lett. {\bf#1},{\ #2} (19#3)}
\def\rmp#1#2#3 {Rev. Mod. Phys. {\bf#1},{\ #2} (19#3)}
\def\zp#1#2#3 {Zeit. Phys. {\bf#1},{\ #2} (19#3)}

\begin{opening}
\title{Recent Results and Perspectives at CDF and D\O\ }

\author{H.E. Montgomery}
\institute{Fermi National Accelerator Laboratory\\
 P.O. Box 500\\
 Batavia, IL60510\
 U.S.A.}

\end{opening}

\begin{document}


\begin{abstract}

 Over the course of the past years the experimental measurements
 performed by the two large collaborations, CDF and D{\O}, at the
 Fermilab Tevatron Collider have fueled advances in our understanding
 of physics at the energy frontier. At the present time the
 accelerator complex and the two detectors are undergoing substantial
 improvements. In this paper, we provide a discussion of some recent
 results which in turn provides a framework within which we can look to
 future prospects.

\end{abstract}

\section{Introduction}

 The Tevatron with the aid of its associated detectors, CDF and D{\O},
 has made its most significant mark on experimental particle physics
 progress with the observation of the top quark. However this was only
 one of a number of important contributions\cite{cdf,d0}. In this
 paper we describe some examples of recent results\cite{mont-moscow98}
 and look forward to future running of the experiments with
 significantly increased luminosity. The latter is possible as a
 result of the introduction of a new accelerator, The Main Injector,
 into the Fermilab complex; this is described in Section 2. Both
 detectors are undergoing upgrades which will enable them to operate
 in the new environment with greatly enhanced capabilities. These
 changes are briefly described in Section 3. In Section 4, we discuss
 the physics accessible at the \ppbar\ collider working our way
 through the physics of the strong interaction, QCD, to the physics
 beyond the standard model. At present much of the latter remains
 speculative but provides the framework for the future experimental
 work. Finally we offer a brief conclusion in Section 5.

\section{The Tevatron and Main Injector}

 The Tevatron is a \ppbar\ collider\cite{mont_dpf99}. The energy is
 900 GeV in each beam and it is the highest energy collider in the
 world. In a hadron collider, the effective parton-parton energy is
 controlled not only by the machine energy but also by the
 luminosity. At high luminosity the rate of higher momentum scatters
 increases. Thus, while \ttbar\ production was accessible
 kinematically from the first day of the collider operation, it was
 not until ten years later that sufficient high momentum scatters had
 been accumulated to make the \ttbar\ production observable
 experimentally.

 The luminosity of the machine is, to a good approximation, controlled
 by the total number of antiprotons available. In turn this depends on
 the production rate and the cooling rate. The new Main Injector, a
 rapid cycling proton synchrotron operating at 150 GeV, will be used
 for antiproton production. An innovation is the introduction of the
 {\it Recycler}, an 8 GeV storage ring constructed with permanent
 magnets. At the end of each Tevatron store, approximately half of the
 initial antiprotons are still present. Most of the luminosity
 degradation is a result of beam blow-up. The antiprotons are to be
 decelerated in the Tevatron, then in the Main Injector to 8 GeV, and
 stored and recooled in the {\it Recycler}. After cooling the the
 anti-proton capacity of the complex is increased by a factor of
 two. With these measures, the instantaneous luminosity in the
 collider is anticipated to increase to about $2~\times 10^{32}$
 {\lumunits}. This will make an integrated luminosity of about 4
 fb$^{-1}$ available in the next few years perhaps rising to 10--30
 fb$^{-1}$ over the next six to seven years. This is to be compared
 with the 0.1 fb$^{-1}$ of the present data set. The Main Injector has
 been comissioned and the Tevatron is operational for fixed target
 physics. At the same time the energy per beam will be increased from
 900 GeV to close to 1000 GeV. For high mass processes, such as top
 production the cross section increases by 30-40\%.

 \section{CDF and D\O\ Detectors}

\begin{figure}[htb]	
\vskip 0.2 cm
\centerline{\hspace{3cm}\epsfxsize 4.5 truein \epsfbox{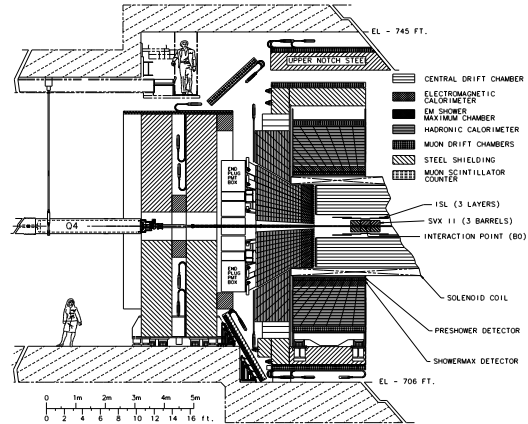}}   
\vskip 0.5 cm
\caption{ The upgraded CDF detector. }
\label{fig-cdf}
\end{figure}

\begin{figure}[htb]	
\centerline{\hspace{1.5in}\epsfxsize 4.5 truein \epsfbox{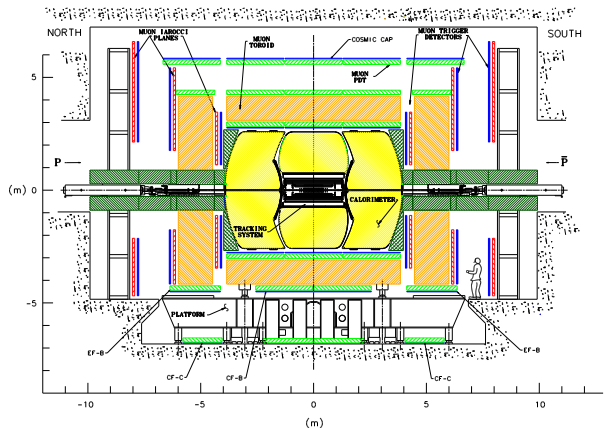}}   
\vskip -2 cm
\caption{The upgraded D\O\ apparatus. }
\label{fig-d0}
\end{figure}

 The CDF detector, see Fig.~\ref{fig-cdf}, has been operational for
 more than ten years. It contains a large solenoid which provides a
 1.4 Tesla magnetic field in the tracking volume. That volume is
 surrounded by a scintillator/lead/iron calorimeter and a non-magnetic
 muon detector.  The current upgrade concentrates on completely
 replacing the tracking detectors. The outer tracking will be provided
 by a large open cell drift chamber with shorter drift distances than
 the previous detector. At inner radii, this is complemented by a
 comprehensive silicon detector system both to enhance the tracking
 capability and to provide detection of secondary decay vertices. The
 detection of $B$ hadrons, both for their own sake and as indicators
 of the decays of higher mass states, top and perhaps the Higgs
 particle, places a strong premium on this capability.

 The D\O\ detector, see Fig.~\ref{fig-d0}, is characterised by a
 three-cryostat liquid Argon/Uranium calorimeter with good electron
 and jet resolutions. The muon system consists of detectors inside and
 outside of large iron toroids in both central and forward
 regions. The forward muon system is being equipped with new trigger
 and tracking detectors to accomodate the upgraded accelerator
 parameters. A new superconducting solenoid has been installed in the
 tracking volume and the particle detection will be performed using a
 scintillating fiber tracker and a 800,000 channel silicon tracker.

 Initially the collider will operate with 496 nsec between collisions
 of the bunches but this will eventually be reduced to 132 nsec.
 Pipelines, analogue in some cases, digital in others, have been
 introduced in the front end electronics in both of the new
 detectors. The data acquisition systems have also been upgraded to
 accomodate event rates of several tens of Hz, to accomodate the
 overall luminosity increase. The detectors are to be operational in
 early 2001.

\section{Physics}

\subsection{QCD}

 \begin{figure}[htb]
\centerline{\epsfxsize 3.0truein \epsfbox{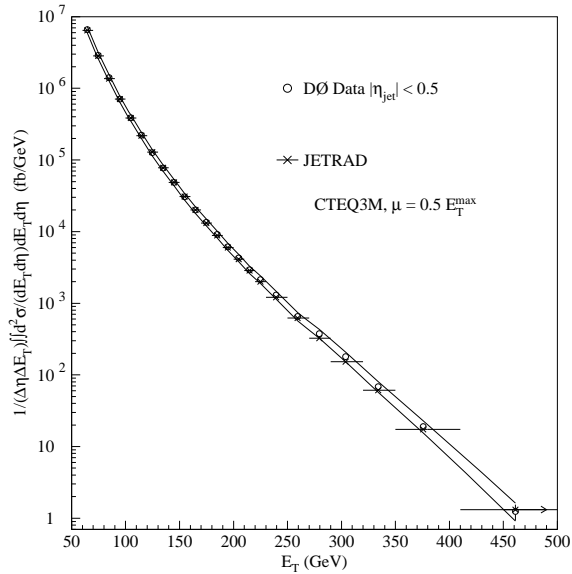}}
\caption{ D\O\ inclusive jets cross section and a comparison with a next-to-leading order QCD prediction.}
\label{fig_d0_inclusives}
\end{figure}

 Production cross sections in \ppbar\ collisions are calculated by
 convoluting the parton distribution functions in proton and
 anti-proton, respectively, with the appropriate hard parton-parton
 scattering cross section\cite{mont_gif98}.  The parton distribution
 functions are derived from a number of measurements, primarily from
 lepton scattering, and are evolved to the appropriate hard scattering
 scale. That this paradigm works well is demonstrated in
 Fig.~\ref{fig_d0_inclusives} where we see agreement of the
 calculations with the inclusive jet production cross sections from
 D\O\ over many orders of magnitude.  Deviations from perfection would
 be signs of either changes required to the parton distribution
 functions or, perhaps, a sign of physics beyond the paradigm. While
 there have been alarms, the currently accepted view is that agreement
 between predictions and data is good\cite{blazey-flaugher}. A
 corollary is that from these results, from measurements of the dijet
 mass distributions, and from measurements of the angular
 distributions, a lower limit on the scale of any
 possible compositeness can be set at about 2-3 TeV.

\begin{figure}[htb]
\centerline{\epsfxsize 3.5truein \epsfbox{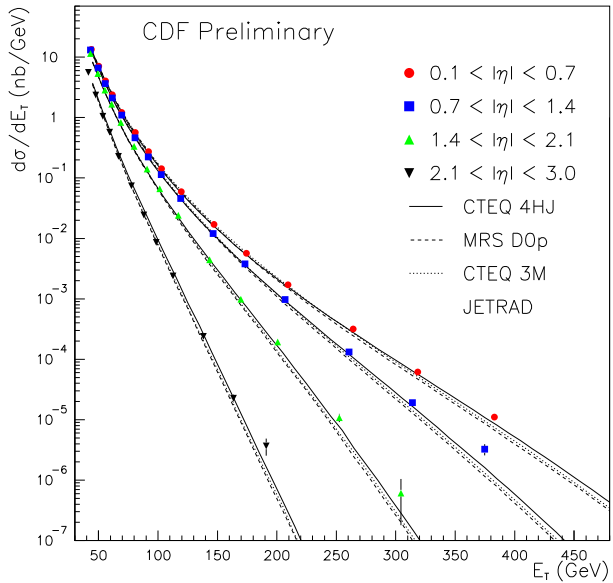}}
\caption{ CDF Inclusive Jets at large rapidity.}
\label{cdf-eta-jets}
\end{figure}

 The prescription described above should also permit an incisive
 comparison between cross sections measured at different energies,
 1800 GeV and 630 GeV, appropriately scaled in transverse
 momentum. This comparison should be fairly insensitive to the choice
 of parton distribution functions; however what is found is that the
 theory differs from the preliminary
 measurement\cite{blazey-flaugher,d0-630gev} by about a factor of
 two. Agreement can only be achieved by modifying the choices of
 renormalisation and factorisation scales.  Recently the measurements
 have been extended\cite{cdf-forwardjets} in rapidity. Again these
 measurements are sensitive to different aspects of the parton
 distribution functions. In this case, as can be seen in
 Fig.~\ref{cdf-eta-jets}, agreement is good.

\begin{figure}[htb]
\centerline{\epsfxsize 3.5truein \epsfbox{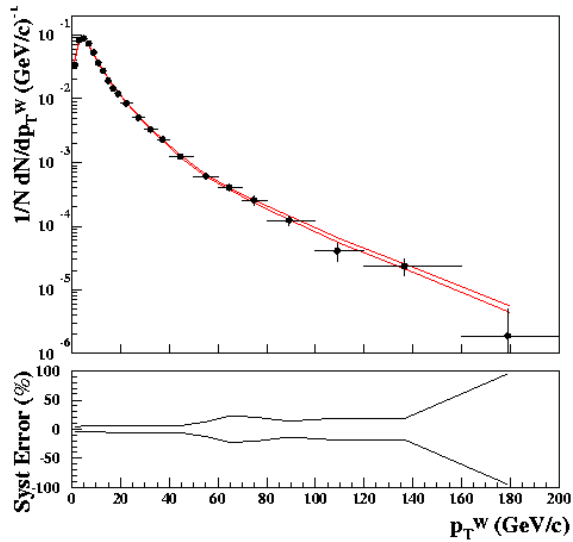}}
\vspace{-1.0in}
\caption{ $W$-boson $p_T$ spectrum as measured by D{\O}, the curves are the bounds of the smeared next-to-leading order predictions. }
\label{d0wpt}
\end{figure}

 Traditionally, experiments at hadron colliders have used a cone
 algorithm\cite{blazey-flaugher} to reconstruct jets. In contrast,
 work using electron-positron, or lepton-nucleon collisions has
 employed algorithms which construct jets from elementary objects such
 as the individual charged tracks or the energy deposited in a
 calorimeter by a single hadron. Such algorithms\cite{ktalg} are known
 as {\it $k_T$}, Jade, or Durham algorithms. Recently we have started
 to use them in \pbarp\ experiments to look at the jet
 substructure. D\O\ has a preliminary result\cite{snihur} which
 suggests that the multiplicity of sub-jets in gluon jets is larger
 than that in quark jets. In the future, this approach could well form
 the basis for distinguishing jet identities in other analyses such as
 Higgs searches.

 When the production of $W$ and $Z$ bosons was first established at
 the CERN S{\pbarp}S there were fewer than ten each of these
 particles. The number of $W$ bosons observed by each of CDF and D\O\
 now approaches one hundred thousand. These data samples have
 permitted the use of the weakly interacting final state particles in
 investigations of QCD, in a manner analogous to the use of virtual
 bosons in the initial state in neutrino experiments.

\begin{figure}[htb]
\centerline{\epsfxsize 3.5truein \epsfbox{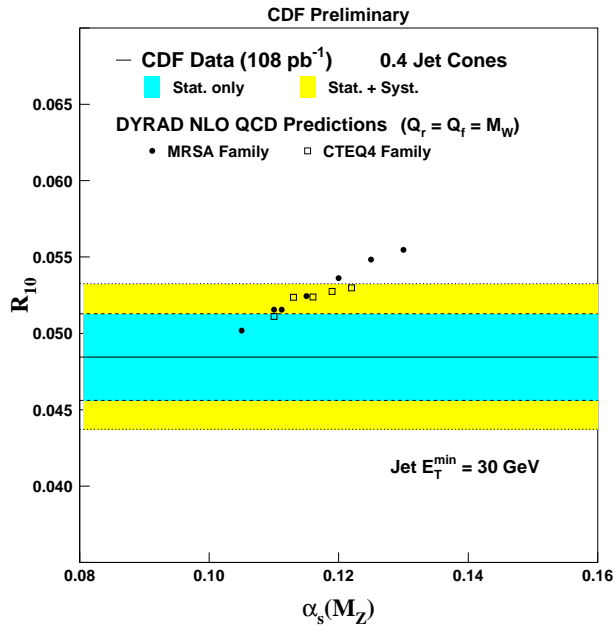}}
\caption{ Fraction of $W$-boson events with at least one jet as measured by CDF. The dependence of the prediction on the strong interaction coupling is also shown. }
\label{cdfr10fig} 
\end{figure}

 $W$ and $Z$ production is dominated by the lowest order parton model
 annihilation of valence quarks from proton and antiproton. The QCD
 calculations agree very well with the cross sections which have been
 measured\cite{wzprod_cs,cdfwzprod} with a precision of a few
 {\%}. The higher order QCD corrections introduce transverse momentum
 for the bosons and at high $p_T$, the recoiling hadronic system may
 contain one or more jets. The transverse momentum spectra at high
 momenta are expected to be well described by perturbative
 calculations while at low $p_T$ non-perturbative effects are
 expected. Resummation of some of the logarithms is expected to be
 necessary. Recent results on both $Z$ and $W$
 spectra\cite{cdfwzprod,d0wptzpt} are surprisingly well described by
 the extant predictions. This is illustrated for $W$ production in
 Fig.~\ref{d0wpt}.

 The CDF measurement\cite{cdfr10} of the fraction of $W$ production
 containing one or more jets above a given threshold, ${\rm{R}}_{10}$,
 is shown in Fig.~\ref{cdfr10fig}. Conceptually, this is a classic
 measurement of the strength of the QCD coupling strength. The events
 with at least a single jet contain at least one strong interaction
 vertex, the total production is dominated by events with no such
 vertex. As usual, since ${\alpha}_{S}$ is not so small, there are
 higher order corrections. Nevertheless, the theory at next to leading
 order provides an adequate description of the results.

\subsection{Flavor Physics}

\begin{figure}[htb]	
\centerline{\epsfxsize 4 truein \epsfbox{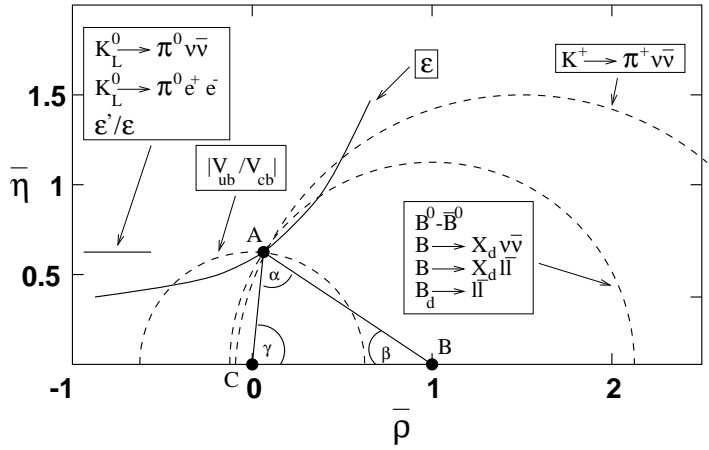}}   
\vskip -.2 cm
\caption{The Unitarity Triangle associated with the Cabbibo-Kobayashi-Maskawa flavor mixing matrix and the measurements possible in the kaon system.}
\label{fig-burasfig}
\end{figure}

 Thus far, three generations of quarks have been observed; each
 generation contains an {\it up}-type quark and a {\it down}-type
 quark. The weak states are mixtures of the eigenstates of the strong
 interaction. This mixing\cite{nakada}, is described by a $3\times{3}$
 matrix of transition amplitudes between the quark states. The
 Cabibbo-Kobayashi-Maskawa matrix can be described by four
 parameters. One of these is a phase through which the formalism
 accomodates and describes the CP violation observed in the kaon
 system. In turn the parameters of the matrix, assuming that there are
 only three generations, can be represented, as shown in
 Fig.~\ref{fig-burasfig}, by a triangle. Many of the properties of
 this triangle are accessible by measurements of the kaon system,
 however measurements of the properties of $B$ hadrons are becoming
 increasingly important.

On the scale of 1800 GeV, the mass of the $b$ quark is small; the
production cross section is about 1/1000 of the total cross
section. Measurements of the cross section by both experiments in the
central region are about a factor of two higher than
expected. Production extends over about six units of rapidity and D\O\
has made measurements\cite{d0fwdmuons} in the forward direction which
are even higher, a factor four, with respect to the predictions.

\begin{figure}[htb]	
\vspace{-0.25in}	
\centerline{\epsfxsize 3.0truein \epsfbox{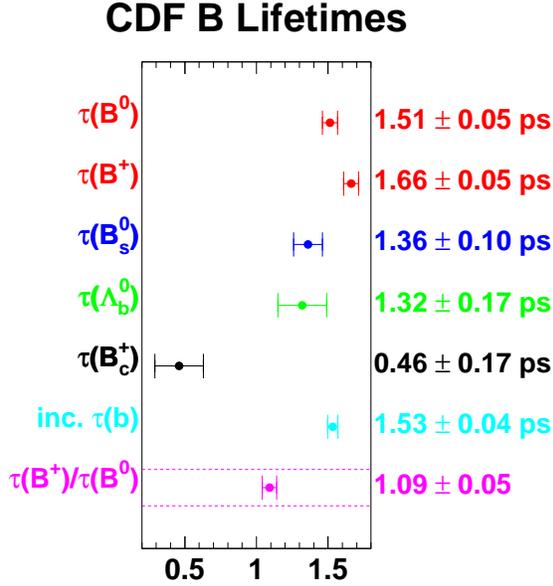}}   
\vskip 0 cm
\caption{A compilation of B hadron lifetime measurements from CDF.}
\label{fig-blifetimes}
\end{figure}

 As a result of the high energy, $B$ hadrons which contain a strange
 quark or a charm quark are produced in addition to those containing
 up and down quarks. For example, in 1998, CDF observed\cite{cdfbc} the
 $B_c$. This state was observed in a semi-leptonic decay mode with a
 missing neutrino. Nevertheless, the signal was unequivocal and a good
 determination of the mass was made.

 Using its silicon vertex detector, CDF has accumulated a set of
 measurements of the lifetimes of various $B$ hadrons. These results
 are displayed in Fig.~\ref{fig-blifetimes}. It is immediately clear
 that, unlike in the charm system, the lifetimes of charged and
 neutral $B$ mesons are rather similar. This is understood to be the
 result of the dominance of the simplest diagrams as a result of the
 high $b$-quark mass. We also notice that the lifetimes of both the
 ${\Lambda}_{b}$ baryon and of the $B_s$ meson are quite similar to
 that of the $B^{+}$ and $B^0$. The lifetime of the $B_c$ is affected
 by the decay of the $c$ quark as well as that of the $b$ quark, hence
 the observed factor of two shorter lifetime\cite{gershteinetal}.

 The neutral $B$ mesons mix in a manner similar to the neutral kaons
 and the mixing for $B_d$ has been measured. Thus far the attempts to
 detect $B_s$ mixing have not met with success. The most recent
 measurements\cite{cdf_b_s} from CDF place a 
 limit(${\Delta}m_s < 5.8$ ps, $x_s = {{\Delta}m_s/{\Gamma_s}} >$ 7.9)
 as good as any achieved by the LEP experiments or by SLD at
 SLAC. Extrapolating to Run II, we expect sensitivities in the range
 $x_s \ge 25$ from each experiment, likely more from CDF, thus
 comfortably covering the expected range.

\begin{figure}[htb]	
\centerline{\epsfxsize 3.5truein \epsfbox{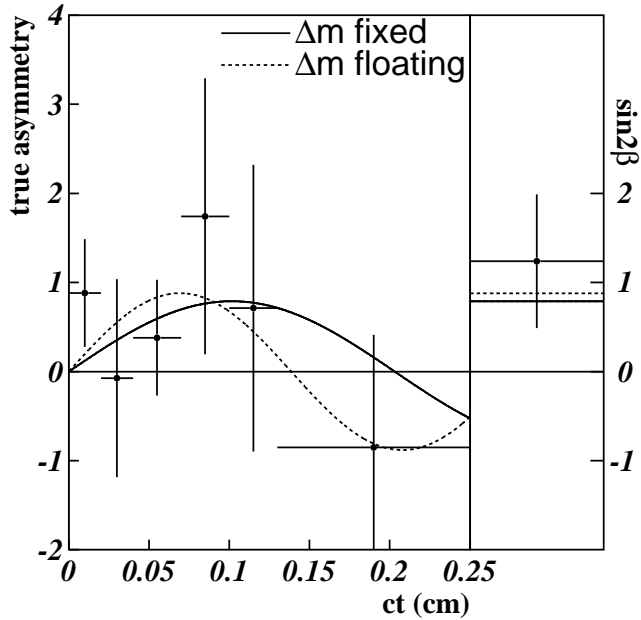}}   
\vskip 0 cm
\caption{Measurement of the CP violating Asymmetry in $J/\psi K^{0}_{S}$ from CDF. On the left the asymmetry is shown as a function of lifetime, and on the right, for an independent sample, the time-integrated asymmetry is shown.  }
\label{cdfsin2beta}
\end{figure}

 The goal of flavor physics experiments is to measure all the
 parameters with sufficient detail to overconstrain the CKM matrix
 and, if possible, to break the model. Recently CDF presented results
 of their measurements of the CP violating asymmetry in the decay $ B
 \rightarrow J/\psi K^{0}_{S}$ which determines $\sin 2\beta$. They
 have one sample of events for which the proper decay time is measured
 and one for which only the time-integrated measurement is
 obtained. The results are displayed in Fig.~\ref{cdfsin2beta}. They
 find\cite{cdf-sin2beta} $\sin 2\beta=0.79_{-0.44}^{+0.41}$ suggesting
 a positive value at about the 90\% c.l.  A feature of this
 measurement is the use of several different flavor tagging
 techniques. With approximately 2~$\rm{fb^{-1}}$ and the upgraded
 detectors, the uncertainty on $\sin 2\beta$ will be reduced below 0.1
 for each experiment. Similar uncertainties are projected for $\sin
 2\alpha$ although the interpretation for this case is considered to
 be more difficult. Measurement of the third angle, $\gamma$, will be
 a challenge.

 The mass of the top quark will be considered as an electroweak
 parameter and discussed in the following section. As far as the
 determination of the couplings and other characteristics of the top
 quark are concerned, studies are in their infancy. The observed cross
 section certainly seems to be consistent with that expected. Further,
 the mix of events of different topologies as yet show little
 deviation from expectations. These facts limit somewhat the liberties
 which can be taken with the coupling $V_{tb}$ between top and bottom
 quarks and possible decays, for example into a charged Higgs boson
 plus a bottom quark.

 One of the fascinating properties of the top quark is its short
 lifetime. This implies that the $W$ boson into which it is expected
 to decay will have a well defined polarisation. CDF has
 measured\cite{cdftopw} that the fraction of longitudinally polarised
 $W$ bosons is $0.97 \pm 0.37(stat) \pm 0.12(syst)$; the expected
 value is 0.70. The short lifetime means that decays occur before
 hadronisation and as a result correlations between the spins of the
 top and anti-top quarks, generated by the annihilation production
 diagram, are expected to survive and manifest themselves in the
 relative spins of the observed states. D\O\ has made a measurement
 using the sample of events in which each of the $W$ bosons decayed
 into a lepton and neutrino. While statistics are small, this sample
 is relatively background free. Along with that of the $b$ quark, the
 angular distribution of the charged lepton is the most powerful
 indicator of this property. The measure of the correlation is the
 parameter $\kappa$ defined such that $ -1 < \kappa < +1$ with $\kappa
 = 1$ expected in the standard model. D\O\ observes\cite{d0topspin}
 $\kappa > -0.25$ at 68\% cl.

\subsection{Electroweak Physics}

 In an earlier section we mentioned the large numbers of $W$ and $Z$
 bosons produced at the Tevatron. In addition to single bosons, boson
 pair production is also kinematically accessible. The various
 possible processes are sensitive to the triple boson couplings. These
 couplings among the gauge bosons are a fundamental feature of the
 non-Abelian electroweak theory and markedly different from the purely
 electromagnetic theory in which the gauge bosons, the photons, carry
 no charge.  In the electroweak theory the non-Abelian couplings lead
 to cancellations among the different diagrams. For example, without
 them the production cross section for several diboson final states
 would diverge at high energy and would violate unitarity. As a
 result, searches for these rare processes have led to
 limits\cite{ellison_review} on the possible deviations from these
 coupling strengths from their standard model values. As an indicator
 of what the future may hold, CDF has observed an excellent candidate
 for the pair production of $Z$ bosons and D\O\ has an excellent
 candidate $WZ$ event.
 
 The copious production of $W$ bosons permits a study of their
 properties. The ratio of production cross section times branching
 ratios for $W$ and $Z$ can be related, using measurements from LEP of
 the $Z$ boson properties, to the total $W$-boson
 width\cite{wzprod_cs}. The standard model value for the $W$ leptonic
 width is used as input as well as the calculated ratio of the
 production cross sections. Alternatively the high-transverse-mass
 tail of the the $W$ boson event distribution can be used to directly
 measure\cite{cdfdirectwidth} the width. At present the indirect
 technique is most precise; however, it does depend on standard
 model assumptions. The hope would be that in the future both
 measurements would be of sufficient precision to provide an
 additional constraint on the standard model and a determination of
 the leptonic width of the $W$ boson.

 The gauge sector of the standard electroweak model is specified by
 three quantities usually taken to be the muon weak decay constant,
 the electromagnetic fine-structure constant and the mass of the $Z$
 boson. These three quantities unambiguously lead to a prediction of
 the mass of the $W$ boson at lowest order in the theory. Higher order
 corrections are expected from loop diagrams containing the fermions
 and hence dominated by the top quark, and a diagram with emission and
 reabsorbtion of the Higgs boson, if such exists. A precise
 measurement of the $W$ boson mass is therefore an important
 goal.

 $W$ bosons are observed by detecting a charged lepton and measuring
 the hadronic recoil vector. This permits the neutrino kinematics to
 be inferred using momentum conservation, but only in the transverse
 plane. The mass of the boson is therefore deduced by fitting
 templates generated with a range of masses to one or all of the two
 lepton transverse momentum spectra and the transverse mass
 spectrum. (The transverse mass is a quantity constructed analogously
 to the effective mass but which is defined using only the transverse
 components of the energy-momentum vectors.) Each of these spectra is
 sensitive to different aspects of the measurement. The charged lepton
 recoils are very sensitive to the $W$ transverse momentum spectrum
 but largely independent of the hadron measurements. The neutrino
 recoil is very sensitive to both the boson transverse momentum and
 the hadronic measurement. The transverse mass is moderately sensitive
 to the lepton and hadronic measurements but relatively insensitive to
 the boson transverse momentum. Hence, use of all the measurement
 information provides powerful cross checks; maximum sensitivity is
 achieved by combining all measurements. These features, in addition
 to the use of the $Z$-boson data for {\it in situ} calibration,
 permit the experiments to match reduction in systematic uncertainties
 to reductions in statistical uncertainties as the samples increase.

\begin{figure}[htb]	
\centerline{\epsfxsize 3.0 truein \epsfbox{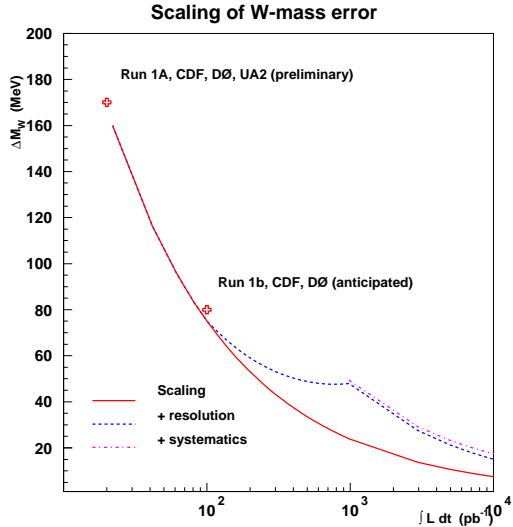}}   
\vskip 0 cm
\caption{ Expected evolution of the precision of a measurement of the $W$-boson mass at the Tevatron Collider. }
\label{fig-wmassevol}
\end{figure}

 The current results $80.433 \pm 0.079$ GeV from the CDF
 measurement\cite{cdfwmass} and $80.482 \pm 0.091$ GeV from the D\O\
 measurement\cite{d0wmass} lead to a combined result of $80.448 \pm
 0.062$. The CDF measurement uses both electrons and muons, primarily
 in the central regions of the detector. The D\O\ measurement is
 limited to electrons but exploits data from the end as well as the
 central calorimeters. These results are in excellent agreement with
 the combined results\cite{lep-ew-wg} of the four LEP experiments of
 $80.350 \pm 0.056$ GeV. They also demand the inclusion of electroweak
 loop corrections to be compatible with the three basic electroweak
 parameters. The good behavior of systematic errors discussed above is
 illustrated in Fig.~\ref{fig-wmassevol} which contains a projection
 of the evolution of the $W$ mass uncertainty as the integrated
 luminosity is increased even further with the upgraded detectors. One
 can note that this projection preceded the most recent results but
 that these 100~pb$^{-1}$ measurements fit well on the curve. The cusp
 occurs as the number of interactions per crossing increases followed
 by a further change in the time spacing between bunch crossings. It
 would seem that an uncertainty of 40 MeV per experiment is not out of
 the question.

\begin{figure}[htb]
\centerline{\epsfxsize 3.5truein \epsfbox{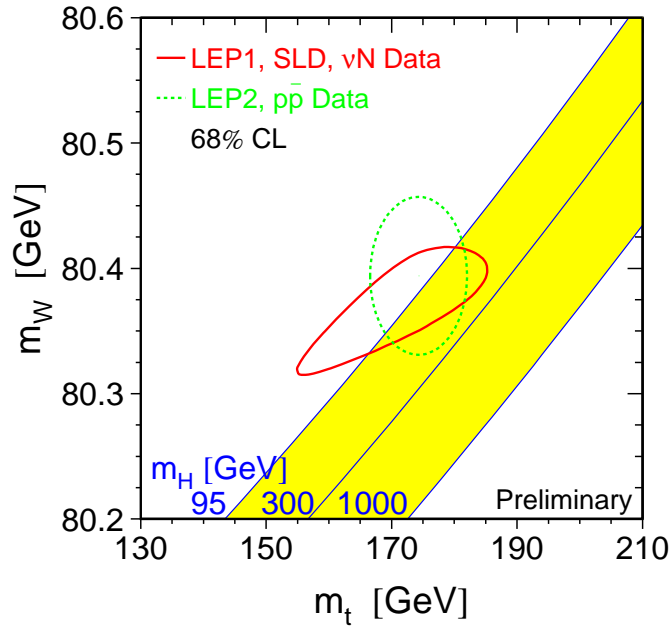}}
\caption{$M_W$ versus $m_t$ showing that a light Higgs is favoured by the current data.}
\label{mwmt} 
\end{figure}

 Because of its large mass, the top-quark is currently only directly
 accesible at the Tevatron. The current data samples have allowed CDF
 and D\O\ to measure\cite{sliwa} the mass. The techniques vary
 depending on the channel used. CDF uses final states in which the top
 and the antitop each decays into three jets, two light quark jets
 from the intermediate $W$ boson and a $b$- or {\bbar}- quark
 jet. However in this channel the background is large and the
 resulting measurement has an uncertainty of around 10 GeV. In the
 dilepton channel, even with two missing neutrinos, the mass can be
 determined; however, the low statisistics offset a rather good
 understanding of the systematic uncertainties. Again from each
 experiment the uncertainties are of order 10 GeV. For each experiment
 the dominant measurement comes from the ``lepton-plus-jets'' channel
 in which one of the intermediate $W$ bosons decays leptonically
 giving a charged lepton and a missing neutrino while the other decays
 into light quarks. The final state then contains four jets and a
 lepton for which each of the momentum vectors is fully measured, and
 a neutrino for which only the transverse components are
 measured. Using the mass constraints, those from the intermediate $W$
 bosons and that from demanding that the top and antitop masses be
 the same, leads to kinematic fits with two constraints. Account has
 to be taken of the possible combinations. These may be restricted if
 the $b$-quark jet is well identified, either by a displaced decay
 vertex or from a soft charged lepton from the $b$-quark decay. Using
 this channel, CDF achieves an uncertainty of about 7 GeV including
 systematics whereas that from D\O\ is 8 GeV. All these measurements
 are combined taking into account the correlations in
 uncertainties, both those derived from common experimental errors
 (between channels in a given experiment, and those derived from
 common techniques between experiments. The result is that $m_{t} =
 174.3 \pm 3.2 \pm 4.0$ GeV. This makes the mass of the top quark the
 best measured of all the quark masses.

 We can then use the top mass and the $W$ mass and compare them with
 the basic electroweak predictions. As indicated earlier, the Higgs
 boson mass enters into the calculations of the electroweak
 loops. Hence the combination of all the electroweak measurements has
 sensitivity to the mass of the putative Higgs boson.This is
 illustrated in Fig.~\ref{mwmt}; we see that the data favour a light
 Higgs, of order 100-200 GeV. The uncertainties are rather large but
 this tendency in the existing data both from the Tevatron and from
 LEP/SLD has lent excitement to the searches current at LEP and to
 work on the upgrades of the Tevatron experiments.

 The top mass determination described above is dominated by the
 uncertainties in the jet energy scale calibration. In future runs we
 expect that these uncertainties can be reduced using the data
 themselves. CDF has observed a $W$-mass peak in the decay jets from
 top and, in a \bbbar\ data sample, has observed the peak from the $Z$
 boson. Taking into account secondary vertex triggers, which each
 experiment expects to use in the upcoming running, the latter will
 provide a powerful jet calibration tool. As a result we can look
 forward to a reduction of the uncertainty on the mass of the
 top-quark to less than 2 GeV from each of CDF and D\O\ in the next
 few years.

\subsection{Beyond the Standard Model}

 It is the duty of experimenters at the highest energy colliders to
 search for phenomena not previously observed. These searches are
 necessarily guided by how we imagine the underlying physics. At
 any given time, certain scenarios enjoy more popularity than
 others. We have seen in the past that states decaying into lepton
 pairs have often provided discoveries, for example the $J/\psi$ and
 the $Z$ boson. QCD breaks the electroweak symmetry, however not with
 sufficient strength to explain the masses of the $W$ and $Z$, so some
 have postulated an analogous interaction, technicolor, to generate
 such masses. At the present time, many theorists believe that
 supersymmetry should play a role and the phenomenology of many
 possible scenarios is well developed. As experimentalists, we keep an
 open mind.

 Searches have been performed for the leptonic decays of higher mass
 gauge bosons both $W$-like and $Z$-like. None have been found with
 masses less than about 600-700 GeV. Similarly, by looking in the mass
 spectra of jet pairs, excited quarks can be excluded with masses less
 than about 600 GeV. However, such states can also influence angular
 distributions of lepton or quark pairs at energies well below their
 masses. Thus such measurements provide windows to very high masses. A
 recent example is the measurement\cite{d0_drellyan} of lepton pair
 production which sets limits of 3-6 TeV on such compositeness
 scales. A few years ago, an excess of high mass events was
 observed\cite{hera} by the HERA experiments. One possible explanation
 was the existence of 1st generation leptoquarks, composite
 electron-quark states. Extensive measurements\cite{hagopian} at the
 Tevatron now exclude such particles unless their masses exceed about
 220 GeV. Similar, if slightly weaker, limits have been place on the
 masses of 2nd and 3rd generation leptoquarks.

\begin{figure}[htb]	
\centerline{\epsfxsize 3.3 truein \epsfbox{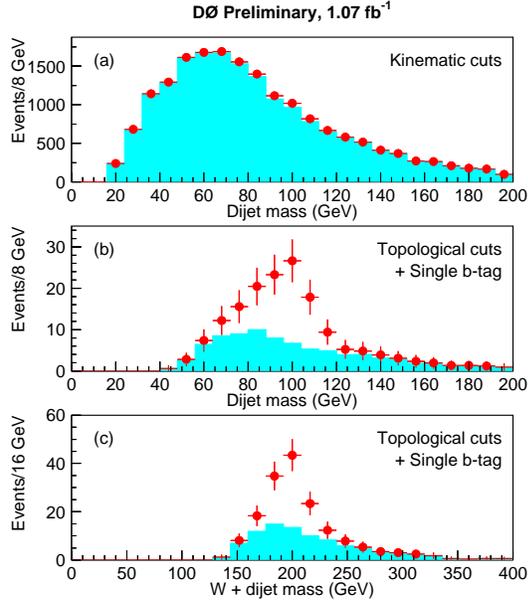}  }   
\vskip 0 cm
\caption{Mass spectra expected for technicolor signals for the $\pi_{T}$, with a mass of 110 GeV, and the $\rho_{T}$, with a mass of 210 GeV. The points are the expected distributions including the technicolor signal; the shaded histogram is the standard model background.}
\label{fig-technicolor}
\end{figure}

 Technicolor searches are relative newcomers to the Tevatron analysis
 menu. Searches have been performed\cite{cdf_technicolor} which
 provide lower limits on the masses of technipions($\pi_{T}$),
 technirho($\rho_{T}$) and techniomega($\omega_{T}$) mesons. The
 techniques are illustrated by Fig.~\ref{fig-technicolor} in which the
 submass distributions of the cascade decay of a technirho( $\rho_{T}
 \rightarrow \pi_{T} + W \rightarrow {b} + \overline{b} + W $) are
 plotted. With the 1 $\rm{fb^{-1}}$ used in this study\cite{womers}
 the signals at 110 and 210 GeV are clearly visible.

\begin{table}[htb]
 \caption{Mass ranges covered for a $5\sigma$ discovery in SUSY models.}
 \begin{tabular}{l|c|cc} 
 Model & SUSY Particle & Run I( 0.1 $fb^{-1}$ &  Run II(2.0 $fb^{-1}$\\
       &               & Mass Limit(GeV)      &    Mass Limit(GeV) \\
 \hline 
 SUGRA &&\\
  \hline
 & $\tilde{\chi}_{1}^{\pm}$ & 70  & 210 \\
 & $\tilde{g}$ & 270 & 390 \\
 & $\tilde{t}_{1}(\rightarrow b\tilde{\chi}_{1}^{\pm})$& & 170 \\
  \hline
 GMSB&&\\
  \hline
 & $\tilde{\chi}_{1}^{\pm}$ & 150 & 265 \\
 & $\tilde{\tau}$ &  & 120 \
 \label{table-susy}
 \end{tabular}
 \end{table}

 The most general Supersymmetry(SUSY) theories have more than one
 hundred parameters and a comprehensive search is almost
 impossible. The usual strategy is to search for signals suggested by
 particular classes of models in which theoretical bias is applied to
 reduce the numbers of parameters to a few.  Searches at the Tevatron
 initially concentrated on the so-called minimal-Supergravity(m-SUGRA)
 models by looking for multi-jet final states with missing transverse
 energy. The missing transverse energy is supposed to be a clear
 signal for the Lightest Supersymmetric Particle(LSP); if R-parity is
 conserved, the LSP survives the decay chain, is neutral, and will
 escape detection since its interactions with the matter of the
 detector are weak. These have been complemented by searches in
 channels containing leptons in the final state. The current
 limits\cite{pagliari} are in the region of squark and gluino masses
 of 270 GeV if the two are approximately equal and squark masses less
 than about 150 GeV are excluded for any gluino mass.

 If R-parity conservation is not assumed, missing transverse energy is
 no longer a useful discriminant. Nevertheless searches have been
 performed\cite{cdf_rpv,d0_rpv} particularly in channels with leptons
 in the final state. Limits on the squark masses in such analyses are
 about 260-280 GeV, and similarly for the gluino when the masses are
 equal.

 In recent years alternative SUSY breaking scenarios to the SUGRA
 models have been explored. In particular, the
 observation\cite{cdf_event} of a single spectacular
 e``e''$\gamma\gamma$\met event prompted considerable
 activity. Theorists attempted to accomodate the event using Gauge
 Mediated Symmetry Breaking(GMSB) which in general leads to numerous
 electromagnetic objects, photons and electrons, in the final state
 along with missing transverse energy. Within the phase space of these
 models, limits on the mass of the lightest chargino, of 120 GeV from
 CDF\cite{cdf_event} and of 150 GeV from D\O\cite{d0_gmsb}, have been
 reported.

 Looking to the future, a summary of what we can expect with about 2
 $\rm fb^{-1}$ is displayed in Table~\ref{table-susy}. These are the
 results of a fairly extensive study\cite{runiihiggssusy} of the potential
 for future SUSY exploration at the Tevatron.

\subsection{The Higgs }

 Almost independently of theoretical religion, it is believed that
 some spin 0 high mass structure, known generically as the Higgs
 boson, must exist to generate the masses of the $W$ and $Z$
 bosons. In its simplest form it could have just one observable
 state. Analysis of the current electroweak data suggest, as we have
 seen, that it be relatively light. If so it could potentially be
 discovered in the current running of LEP II\cite{fernandez}, or, as
 concerns us here, at future running of the Tevatron.

\begin{figure}[htb]	
\vspace{-0.5cm}
\centerline{\epsfxsize 1.728 truein \epsfbox{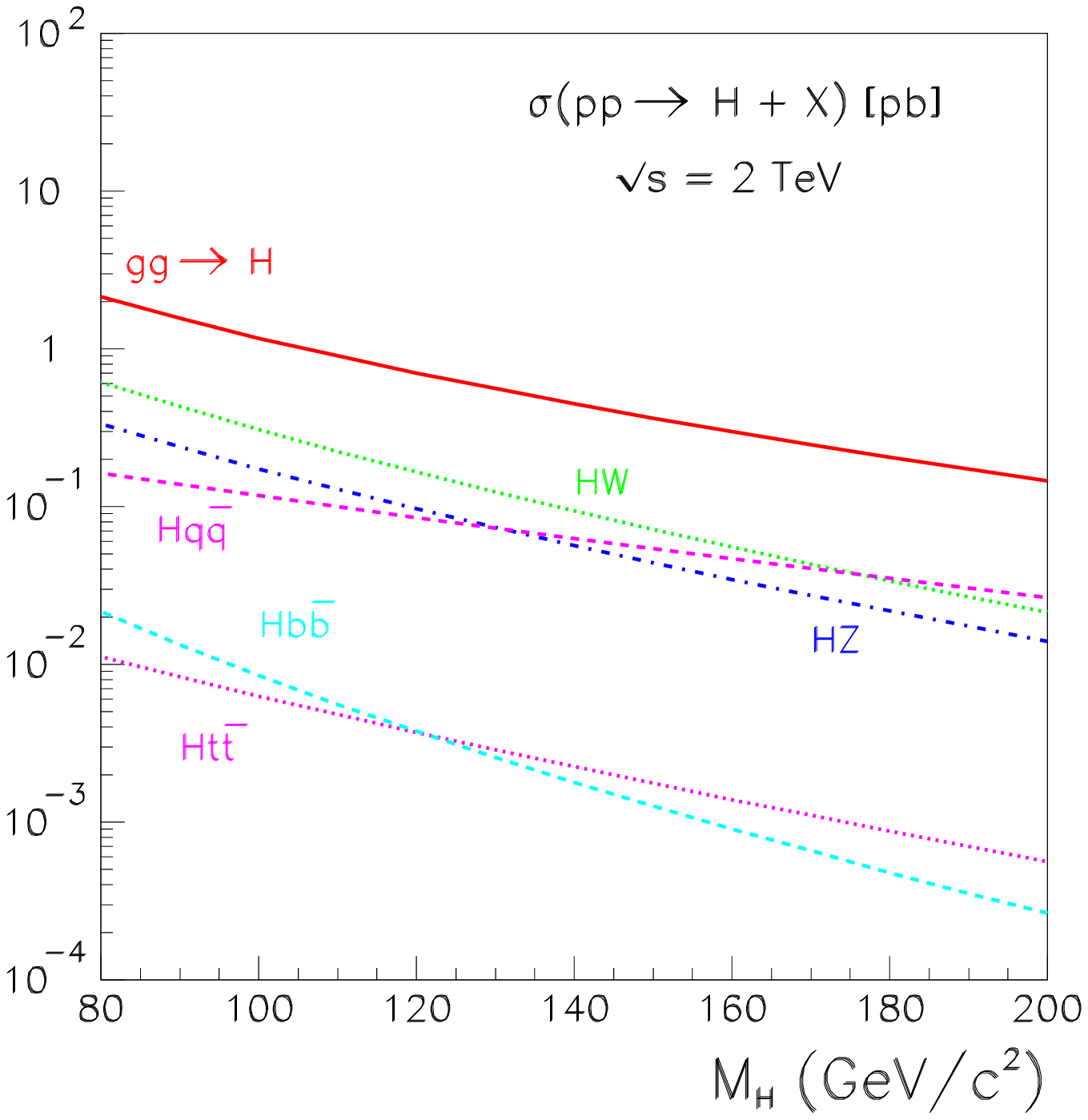} \epsfxsize 1.568 truein \epsfbox{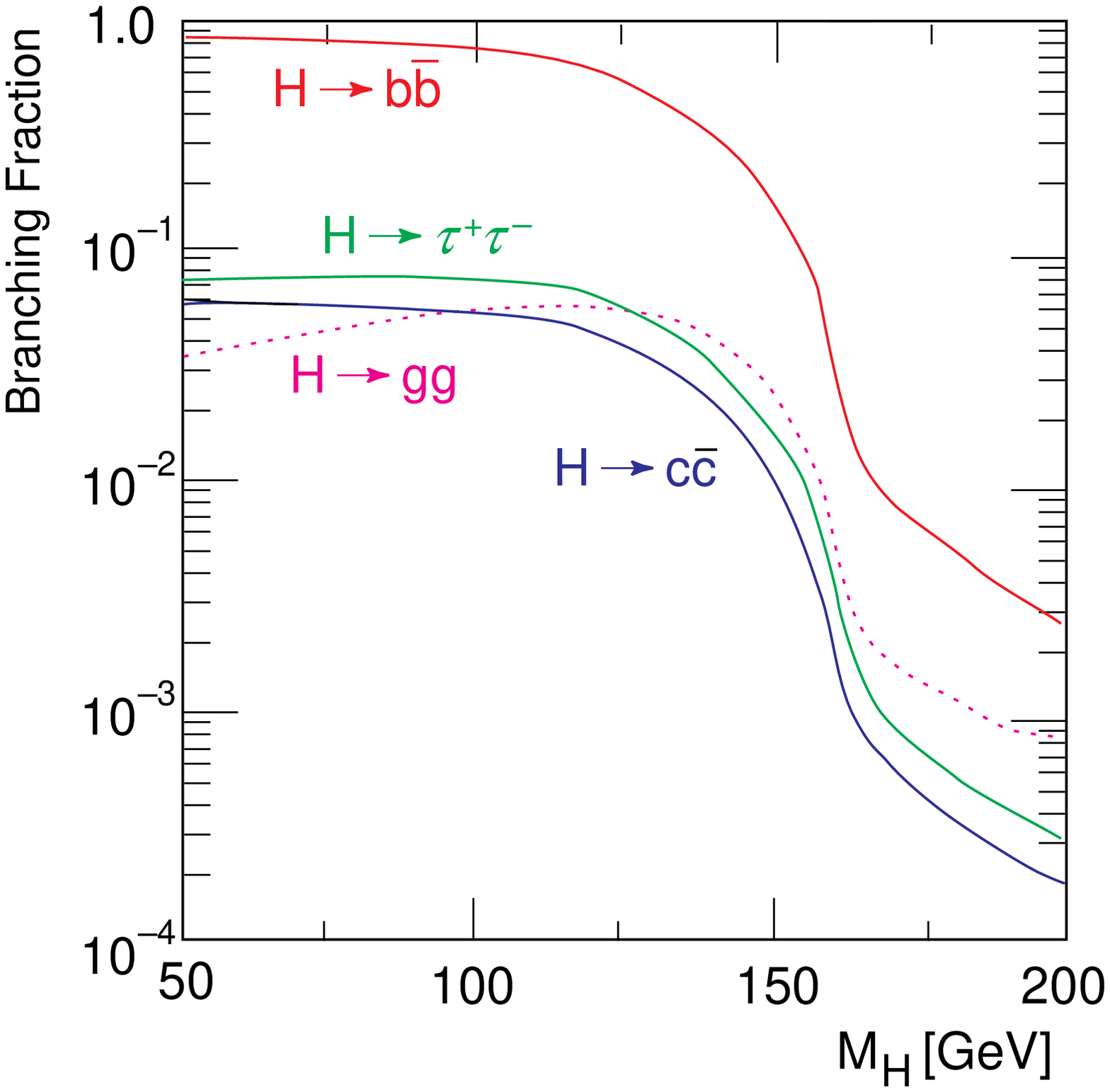}\epsfxsize 1.568 truein \epsfbox{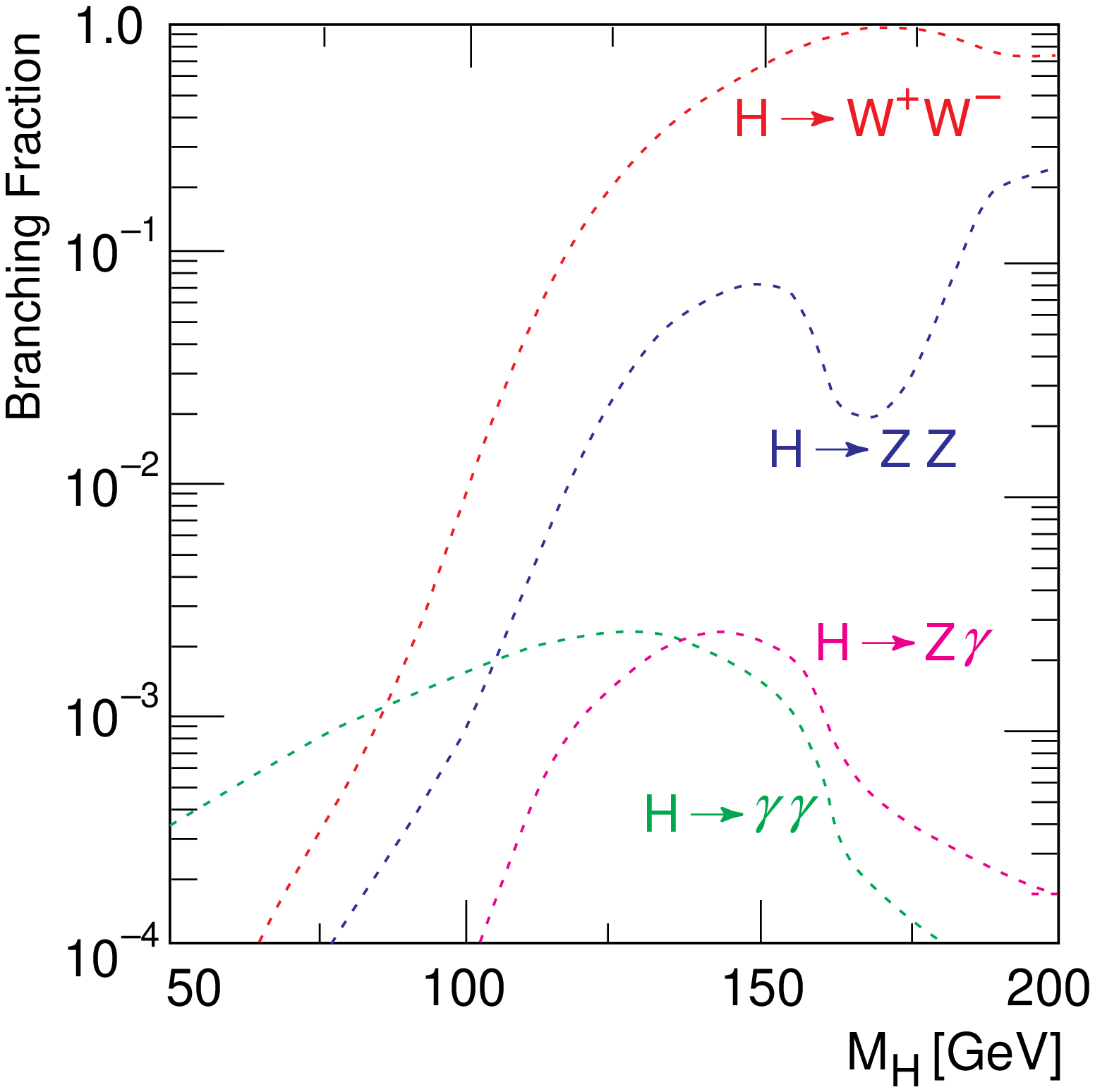} }   
\vskip 0 cm
\caption{Higgs-boson production cross sections and  branching fractions to fermions and to bosons as a function of Higgs-boson mass.}
\label{fig-higgsbr}
\end{figure}

 The standard model Higgs particle may be produced at the Tevatron in
 several different ways. As indicated in Fig.~\ref{fig-higgsbr}, the
 cross section for the gluon fusion process is about 1 pb at 100
 GeV. The ``associated production'' of an electroweak gauge boson, $W$
 or $Z$, and a Higgs boson is approximately an order of magnitude
 less. The decay branching ratios to fermions and to bosons are also
 shown in the other two diagrams in Fig.~\ref{fig-higgsbr}. As
 expected, at low masses the \bbbar\ mode dominates. However one
 notices that the $WW$ mode becomes large for masses in excess of 130
 GeV. Folding these facts together and taking into account the need
 for a distinctive signature on which to trigger, and on which to key
 the analysis, two distinct approaches are discussed.

\begin{figure}[htb]	
\centerline{\epsfxsize 4.0truein \epsfbox{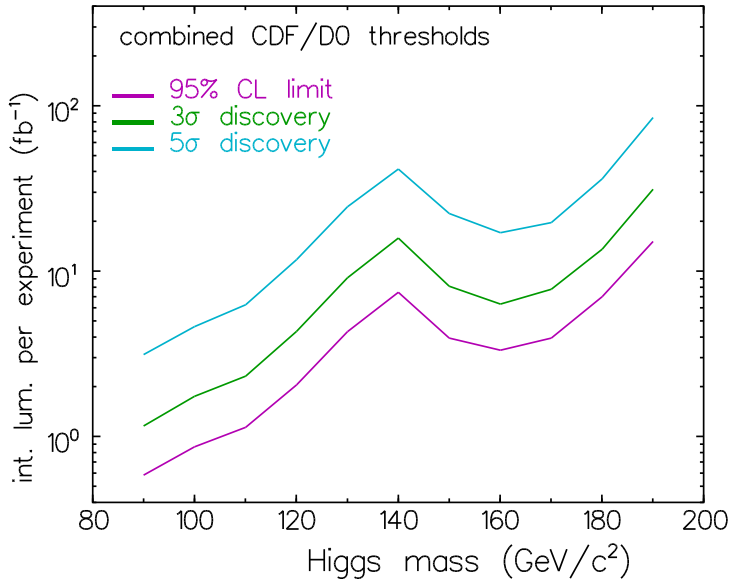}}   
\vskip 0 cm
\caption{Luminosity required as a function of Higgs mass to achieve different levels of sensitivity to the standard-model Higgs boson.From the upper curve corresponds to a $5~\sigma$ discovery, the middle a $3~\sigma$ signal and the lower a 95\% exclusion limit. These limits require two experiments, Bayesian statistics are used to combine the channels and   include the improved sensitivity which would come from multivariate analysis techniques.}
\label{fig-higgsensitivity}
\end{figure}

 At low masses the approach is to look for a signal in associated
 production exploiting both the decays of the $W$ or $Z$ and the \bbbar\
 or $\gamma\gamma$ decay modes of the Higgs. 

 Current searches for a bosophilic Higgs use two
 jets\cite{d0_bosophilic}, or two leptons\cite{cdf_bosophilic} from
 the $W$ and $Z$ decays, along with two photons from the Higgs decay.
 D\O\ and CDF respectively place lower mass limits in the region of 80
 GeV.  Using the \bbbar\ mode as a signal for the Higgs, and including
 the $\nu\overline{\nu}$ decay among the lepton decays of the Z,
 limits relevant to the standard-model Higgs are
 obtained\cite{cdf_sm_higgs,cdf_bosophilic}. These do not give a mass
 limit but constrain the cross section at about ten times its standard
 model expected value. This illustrates the premium on integrated
 luminosity. In SUSY models the Higgs structure is more complicated
 and a charged Higgs exists. If its mass would be less than that of
 the top quark then the $H^{\pm}b$ decay mode would compete with that
 of the top quark. The extent to which this would occur is controlled
 by the $\tan\beta$ SUSY parameter. The agreement between the observed
 cross section times branching ratio for the top quark into modes
 containing a $W$ boson and the theoretical prediction has therefore
 been used to place limits\cite{cdf_charged_higgs,d0_charged_higgs} in
 the space of $\tan\beta$ and $m_H$.

 With the upgraded detectors, the efficiencies are improved by a
 factor of about four as a result of the improved detection of $b$
 quarks, and, perhaps key, the ability of each experiment to trigger
 on displaced vertices. Recent studies\cite{runiihiggssusy} take into
 account these improvements and improved understanding of the jet
 calibrations and resolutions. They find that in the relatively low
 mass region, below about 130 GeV, the prospects for observation of a
 standard model Higgs are promising. If it turns out to be possible to
 use all of the leptonic and haronic decay modes of $W$ and $Z$ along
 with just the \bbbar\ decay mode for the Higgs, exclusion up to 140
 GeV and a three standard deviation hint up to about 130 GeV could be
 obtained with 10 $\rm{{fb}^{-1}}$. For higher masses, the key appears
 to be the use\cite{hanturcot} of the $WW^{*}$ and similar decay
 modes. Life is not easy when multiple modes are necessary for the
 observation, nevertheless, the observation of the top quark was
 considerably strengthened by such techniques. A 95\% exclusion up to
 180 GeV might be achievable with the same 10 $\rm{{fb}^{-1}}$ of
 integrated luminosity. A hint of a signal anywhere up to a mass of
 180 GeV probably requires 20 $\rm{{fb}^{-1}}$ of integrated
 luminosity. This is not excluded and is a challenge which the
 experiments and the accelerator are keen to accept.

\section{Conclusions}

 In these lectures we have attempted to describe the Tevatron Collider
 complex, the experiments CDF and D\O\ currently undergoing major
 upgrades, the physics that has come out of 100 pb$^{-1}$ of
 integrated luminosity and to give a sense of the prospects for the
 future. It is safe to say that much of the physics at the Tevatron
 Collider has been a revelation. The experimental environment has
 proved to be tractable. The events are busy but the objects relevant
 to physics, characterised by transverse momenta in excess of 100 GeV
 and of masses in the range 100 to 500 GeV, are readily observable and
 measurable. In particular it has proved possible to calibrate the
 experiments and the analyses so that precision measurements of the
 $W$ mass, the top quark mass and the jet cross section have been
 completed. Finally, complete $B$ states have been reconstructed, and
 their lifetimes measured and a first measurement of the CP violation
 parameter $\sin{2}\beta$ has been made. If any doubts existed a
 decade ago as to the breadth of the potential of the Tevatron, none
 should exist today.

\section{Acknowledgements}

 In this electronic age we are used to having access at the push of a
 button to descriptions and diagrams produced in many parts of the
 world. This has greatly facilitated the preparation of these
 lectures. I would therefore like to thank my many colleagues on the
 CDF and D\O\ experiments who have helped in this work either
 knowingly or unknowingly. John Ellison, Paul Grannis and Nick Hadley
 were kind enough to read the manuscript and suggest corrections.
 Finally the school was immensely enjoyable and I would like to
 express my appreciation to the organisers, the other speakers and to
 the students from all of whom I learned much.

\end{document}